\documentclass[aps,pra,reprint,showpacs]{revtex4-1}

\pdfoutput=1

\usepackage{graphicx}

\usepackage{amsmath}
\usepackage{bm,hyperref,lmodern,exscale}
\hypersetup{colorlinks=true,linkcolor=blue,citecolor=blue,urlcolor=blue}

\let\originalleft\left
\let\originalright\right
\renewcommand{\left}{\mathopen{}\mathclose\bgroup\originalleft}
\renewcommand{\right}{\aftergroup\egroup\originalright}

\begin{document}

\frenchspacing

\title{Positronium collisions with rare-gas atoms}

\author{G. F. Gribakin}
\author{A. R. Swann}
\affiliation{School of Mathematics and Physics, Queen's University Belfast,
Belfast BT7 1NN, U.K.}
\author{R. S. Wilde}
\affiliation{Department of Natural Sciences, Oregon Institute of Technology, Klamath Falls, 
Oregon 97601, U.S.A.} 
\author{I. I. Fabrikant}
\affiliation{Department of Physics and Astronomy, University of Nebraska,
Lincoln, Nebraska 68588-0299, U.S.A.}

\date{\today}

\begin{abstract}
We calculate elastic scattering of positronium (Ps) by the Xe atom using the 
recently developed pseudopotential method [I.~I.~Fabrikant and G.~F.~Gribakin, \href{http://dx.doi.org/10.1103/PhysRevA.90.052717}{Phys. Rev. A} \textbf{\href{http://dx.doi.org/10.1103/PhysRevA.90.052717}{90}}, \href{http://dx.doi.org/10.1103/PhysRevA.90.052717}{052717} (\href{http://dx.doi.org/10.1103/PhysRevA.90.052717}{2014})] and review general features of Ps
scattering from heavier rare-gas atoms: Ar, Kr, and Xe. The total scattering cross section is dominated by two contributions: elastic scattering and Ps ionization
(breakup). To calculate the Ps ionization cross sections we use the 
binary-encounter 
method for Ps collisions with an atomic target. Our results for the ionization cross section agree well with previous calculations carried out in the impulse approximation. Our total Ps-Xe cross section, when 
plotted as a function of the projectile velocity, exhibits similarity with the electron-Xe cross section for the collision velocities higher than 0.8~a.u., and agrees very well with the measurements at Ps velocities above 0.5~a.u.

\end{abstract}


\maketitle
\section{Introduction}
Recently observed similarities between positronium (Ps) scattering and 
electron
scattering from a number of atoms and molecules \cite{Bra10,Bra10a,Bra12} 
in the intermediate energy range were explained \cite{Fab14a,Fab14b} by the
dominance of the electron exchange interaction with the target atom or
molecule. An explicit proof of this equivalence was given using the framework of the impulse approximation \cite{Fab14a}, valid above the Ps ionization threshold. However, at lower energies the
impulse approximation breaks down and more sophisticated methods are required. The close-coupling method, which includes an
expansion of the total wave function over the states of the target and the
projectile, is very challenging computationally. So far, such calculations have been carried only for simple targets like the hydrogen and helium atoms, often using only a small number of states \cite{Sin00,Gho01,Bla02,Wal04}. 

Recently we developed a pseudopotential method \cite{Fab14b} in which a nonlocal Ps-atom potential is constructed based on the electron-atom and positron-atom scattering phase shifts. This method was successfully applied to the calculation of Ps scattering from Ar and Kr, and gave results in good agreement with those of the beam experiments \cite{Bra10}. 

In the present paper we complete our theory for heavier rare-gas atoms by
performing calculations of Ps scattering from xenon. An interesting aspect of this problem is the question of existence of the Ramsauer-Townsend (RT) minimum in the scattering cross sections. It is well known that the RT minimum does exist in electron scattering from Ar, Kr, and Xe. However, our previous calculations 
\cite{Fab14b} did not find it in Ps-Ar and Ps-Kr scattering. We explained this by the relative weakness of the van der Waals interaction between Ps and a neutral atom as compared to
the polarization interaction in electron-atom or positron-atom scattering. This results in positive scattering lengths for Ps-Ar and Ps-Kr collisions, in contrast to the negative scattering lengths in $e^\pm $-Ar and $e^\pm $-Kr scattering. Moreover, instead of the RT minimum, we obtained what could be called the ``anti-Ramsauer maximum'', due to the fast increase of the $S$- and $P$-wave contributions to the elastic scattering cross section.

The present calculations confirm the above observations for Ps-Xe collisions. 
We also extend the earlier work \cite{Fab14b} by developing a method for the calculation of the ionization cross sections for Ps collisions with rare-gas atoms, based on the binary-encounter approach. The results for the total cross sections agree very well with measurements
above the Ps break-up threshold.
However, recent beam measurements \cite{Bra15} 
of Ps scattering by Ar and Xe show that the cross section decreases towards lower energies,
which is not supported by the present calculations.

\section{P\lowercase{s}-X\lowercase{e} collisions and a summary for the heavier rare-gas atoms} 

The pseudopotential method for the calculation of elastic Ps-atom scattering
was developed and described in detail in Ref. \cite{Fab14b}, and only a brief account is given here. We first calculate the positron-atom and electron-atom scattering phase shifts in the static (for $e^+$) and static-exchange (for $e^-$) approximations. We then construct a local positron-atom pseudopotential and an $l$-dependent electron-atom pseudopotential. The latter contains a repulsive core which allows one to decrease the overlap between the wave function of the scattered electron and the occupied atomic orbitals \cite{Bar74}. Atomic units are used throughout.

\begin{figure}
\includegraphics*[width=\columnwidth]{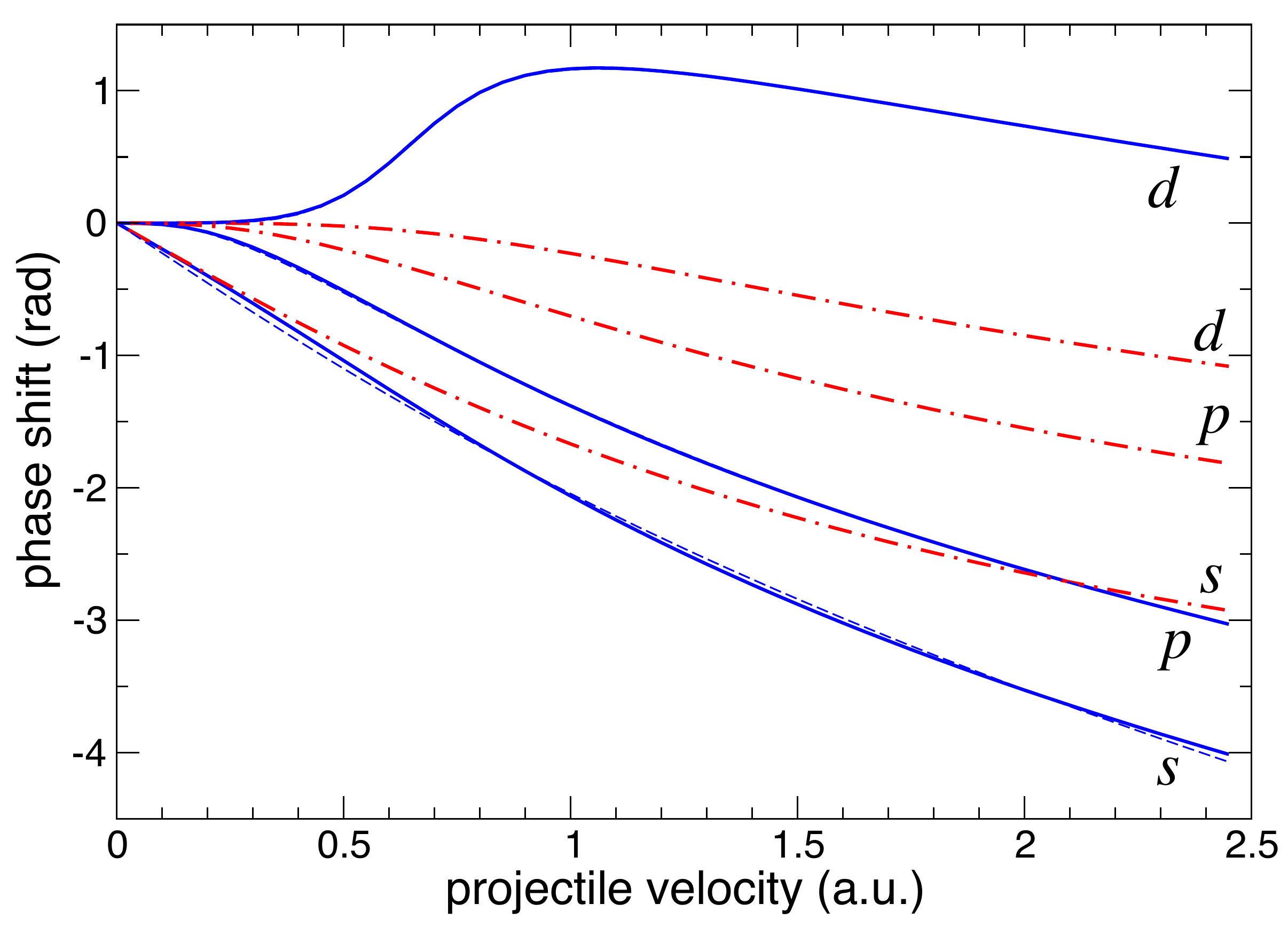}
\caption{Electron and positron $s$-, $p$-, and $d$-wave phase shifts for scattering from Xe calculated for in the static (positron, dot-dashed red lines) and static-exchange (electron, solid blue lines) approximations. The phase shifts obtained using the electron pseudopotential, Eq. (\ref{V_e}), with the parameters given in Table \ref{tb:parameters}, are shown by dashed blue lines. The pseudopotential phase shifts for the electron $d$ wave and all positron partial waves are indistinguishable from the static-exchange/static phase shifts on the scale of the graph.}
\label{fig:el_pos_phases}
\end{figure}

Figure \ref{fig:el_pos_phases} shows the $s$-, $p$-, and $d$-wave phase shifts for positron and electron scattering from the ground-state Xe atom described in the Hartree-Fock approximation. The pseudopotential for the
positron is chosen in the form 
\begin{equation}\label{V_p}
V_p(r)=\frac{Z_p}{r}e^{-\alpha _pr}.
\end{equation}
With the parameters $Z_p$ and $\alpha _p$ chosen as indicated in Table \ref{tb:parameters}, it gives phase shifts that are indistinguishable from the static positron phase shifts on the scale of the plot.

The electron pseudopotential is chosen as
\begin{equation}\label{V_e}
V_e(r)=-\frac{Z_e}{r}e^{-\alpha _er}+\frac{B}{r^n}e^{-\beta r},
\end{equation}
where the second term in Eq.~(\ref{V_e}) represents the repulsive core and exchange. 
The parameters $Z_e$, $\alpha _e$, $B$, $n$, and $\beta $ are adjusted to obtain the best fit of the Hartree-Fock phase shifts (taken modulo $\pi $). In the present calculation we have chosen $n=2$, with all other parameters 
listed in Table \ref{tb:parameters}. Note that the electron 
pseudopotential is $l$ dependent. This is required to effectively 
describe the effect of the Pauli exclusion principle due to the occupied ground-state 
electron orbitals of the Xe atom. The corresponding phase shifts are shown 
in Fig. \ref{fig:el_pos_phases} by thin dashed lines.

\begin{table}
\begin{ruledtabular}
\caption{\label{tab:pseudo}Parameters of the positron-Xe and electron-Xe pseudopotentials.}
\label{tb:parameters}
\begin{tabular}{ccccccc}
projectile & $l$ & $Z_{p,e}$ & $\alpha _{p,e}$ & $B$  & $\beta$ \\
\hline
$e^+$ & 0--4 & 25.09 & 1.568  &  --     &   --\\
\hline
$e^-$ & 0   & 54.00  & 2.8522 & 129.790 & 1.53130 \\
        & 1 & 54.00  & 2.0522 & 50.1760 & 0.88638 \\
        & 2 & 24.713 & 1.0731 & 4.3433  & 0.54959  \\
        & 3 & 15.163 & 1.2381 & $-3.6233$ & 1.0710   \\
        & 4 & 14.792 & 1.3071 & $-3.7086$ & 1.0999   
\end{tabular}
\end{ruledtabular}
\end{table}

After averaging the sum of the positron and electron pseudopotentials over the electron charge distribution in Ps($1s$), we obtain a nonlocal central potential which describes the Ps-Xe interaction in the static approximation. We then add the van der Waals interaction in the form
\begin{equation}\label{vdW}
V_W(R)=-\frac{C_6}{R^6}\left\{1-\exp\left[-\left(\frac{R}{R_c}\right)^8\right]\right\},
\end{equation}
where $C_6$ is the van der Waals constant and $R_c$ is a cutoff radius.
The $C_6$ constant is calculated using the London formula \cite{Lon37}, $C_6=240.6$~a.u., which is accurate to 5\% \cite{Swa15}.
In previous calculations the cutoff parameter $R_c$ was varied between 2.5 and 3.0 a.u. with insignificant change in the results for cross sections.
In the present calculations for Xe we chose $R_c=3$ a.u.

\begin{figure}
\includegraphics[width=\columnwidth]{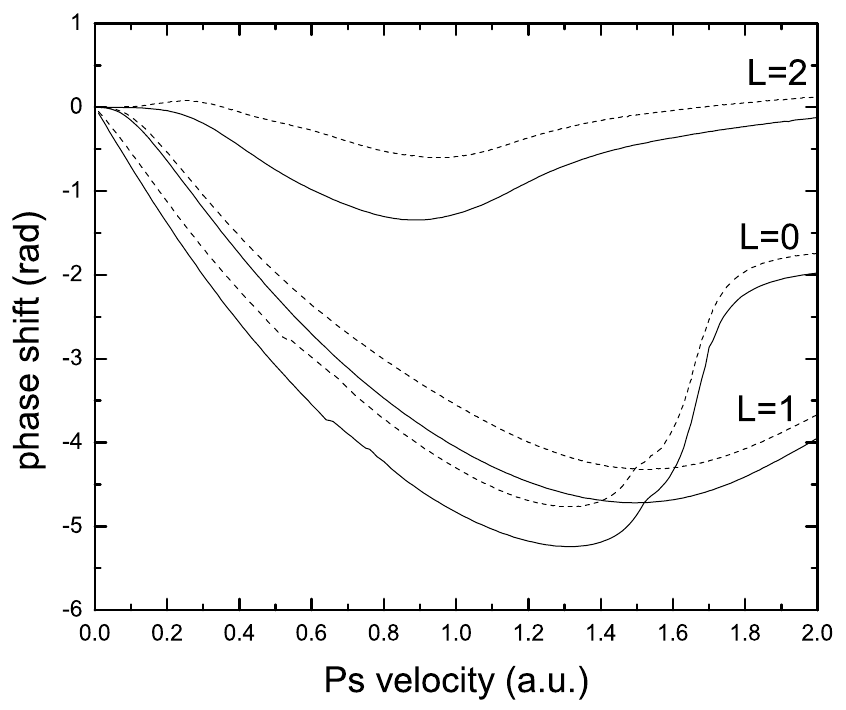}
\caption{Elastic scattering phase shifts for Ps on Xe, obtained with the 
static-field pseudopotential (solid lines), and with the van der Waals 
potential added (dashed lines).}
\label{fig:Xe_ph}
\end{figure}

The integro-differential radial equation for the wave function of the Ps centre-of-mass motion (Eq.~(16) in Ref. \cite{Fab14b}) is solved iteratively. With a suitable choice of the local part of the interaction potential, this process converges quickly, and the solutions yield the Ps-atom scattering phase shifts. These are shown in Fig. \ref{fig:Xe_ph} for the three lowest Ps partial waves, $L=0$, 1, and 2. As in the case of Ar and Kr \cite{Fab14b}, inclusion of the van der Waals attraction gives a positive contribution to the scattering phase shifts. For the $S$ and $P$ waves this leads to a decrease in the scattering cross sections. This effect of virtual excitations of the target and projectile (which
underpins the van der Waals interaction) was discussed previously in Ref. \cite{Lar12}. Similarly, the scattering length for Ps-Xe scattering obtained with the van der Waals potential, $A=2.45$~a.u., is smaller than the value $A=3.57$~a.u. obtained in the static approximation. The latter value is in a reasonable agreement with $A=3.77$~a.u. obtained by Blackwood {\it et al.} \cite{Bla02a,Bla02b} in the static-exchange approximation.
We see that the Ps scattering length for Xe is greater than those for Ar and Kr, which confirms our prediction \cite{Fab14b} of the growth of the positive scattering length with atomic number $Z$. Although the van der Waals interaction (which makes $A$ smaller) increases with $Z$, the effect of the Pauli repulsion for heavier atoms is stronger. Mitroy and Bromley \cite{MB03} used the stochastic variational method with model polarization potentials for the electron- and positron-atom interactions, and obtained values of the Ps-Xe scattering length in the range 1.50--2.60 a.u., with the recommended value of 2.29~a.u., in close agreement with our value.

Turning to higher energies, we observe an unusual increase of the phase shifts
at $v\gtrsim 1.3$ a.u. The $S$-wave phase shift also exhibits a small kink between 
$v=1.5$ and 1.6 a.u. This behaviour reflects the influence
of the repulsive wall in the pseudopotential, and in fact the well-known 
failure \cite{Bar74} of
the pseudopotential approach at higher energies. However, the $S$-wave
and $D$-wave contributions are small in this energy region (${\gtrsim}60$ eV)
and do not influence the behaviour of the total cross section.
  
\begin{figure}
\includegraphics[width=\columnwidth]{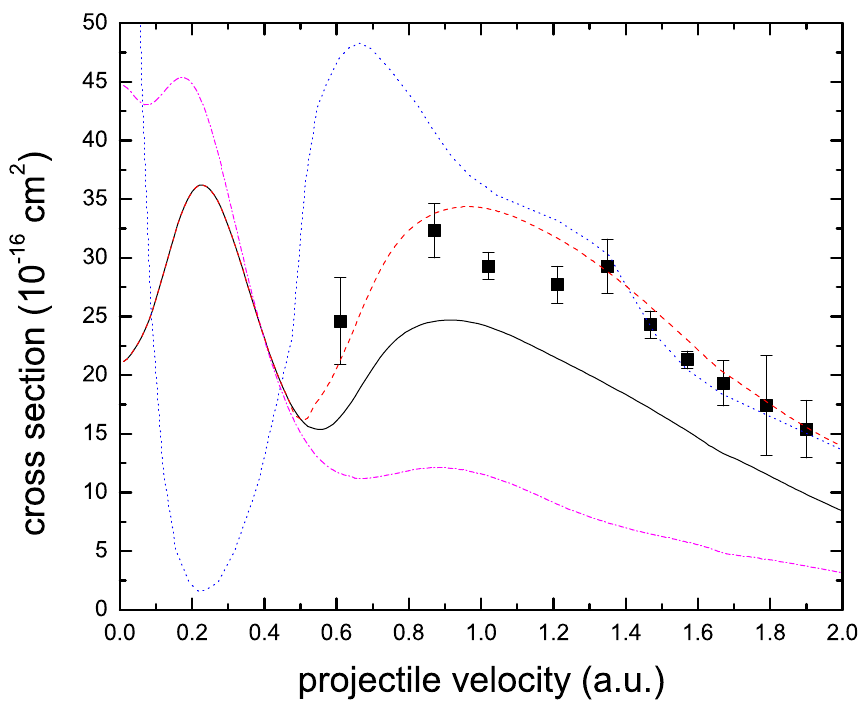}
\caption{Cross sections for Ps and electron collisions with
Xe atoms as functions of the projectile velocity: 
solid black line, present elastic Ps-Xe cross section; dashed-dotted 
magenta line, elastic cross section obtained in the static approximation (i.e.,
without the van der Waals interaction);
dashed red line, total Ps-Xe cross section; dotted blue line, $e^-$-Xe elastic 
scattering cross section taken from the calculations \cite{Sin82} and 
measurements \cite{Dab80}; solid squares, measured Ps-Xe total cross section 
\cite{Bra12}.
}
\label{fig:Ps-Xe}
\end{figure}

The cross sections for Ps-Xe scattering are shown in Fig. \ref{fig:Ps-Xe}. 
In the low-velocity range, the elastic cross section in the static 
approximation (i.e., not including the van der Waals interaction) is close 
to the corrected static-exchange results of Blackwood {\it et al.} \cite{Bla02a,Bla02b} (not shown on the graph). However, at higher velocities, in the range $v=0.4$--1.0~a.u., our cross section decreases more rapidly and is substantially lower than that of Blackwood {\it et al.} \cite{Bla02a,Bla02b}.

The present total scattering cross section is shown in Fig. \ref{fig:Ps-Xe} 
by the dashed red line. It was obtained by adding the elastic and ionization cross sections, the latter calculated using the binary-encounter method (see Sec. \ref{sec:ioniz}). As was shown 
before \cite{Fab14a,Bla02a}, elastic scattering and Ps ionization are the two major processes contributing to the total scattering cross section.

In Fig. \ref{fig:Ps-Xe} we also compare the total scattering cross section with the
measurements and the $e^-$-Xe cross section. The agreement with the experimental data 
\cite{Bra12} is remarkably good. Note, however, that recent measurements below the Ps ionization threshold \cite{Bra15} indicate that the cross section continues to decrease towards 
lower velocities, in contrast with our prediction of
a maximum in this energy region. 
To analyze the similarity between $e^-$-Xe and Ps-Xe scattering, we also 
present the $e^-$-Xe total cross section calculated by Sin Fai Lam \cite{Sin82} for $E<30$ eV ($v<1.485$ a.u.). At higher velocities we show the measured cross section from Ref.~\cite{Dab80}, which agrees very well with the calculations \cite{Sin82} below 30~eV. 
The Ps-Xe cross section 
remains substantially lower than the corresponding $e^-$-Xe cross section 
 for velocities up to 1 a.u. 
This makes the Xe case somewhat different from those of Ar and Kr, where 
the proximity of the electron and Ps scattering cross sections was observed 
right from the ionization threshold $v=0.5$~a.u.

\begin{figure}[ht!]
\includegraphics[width=\columnwidth]{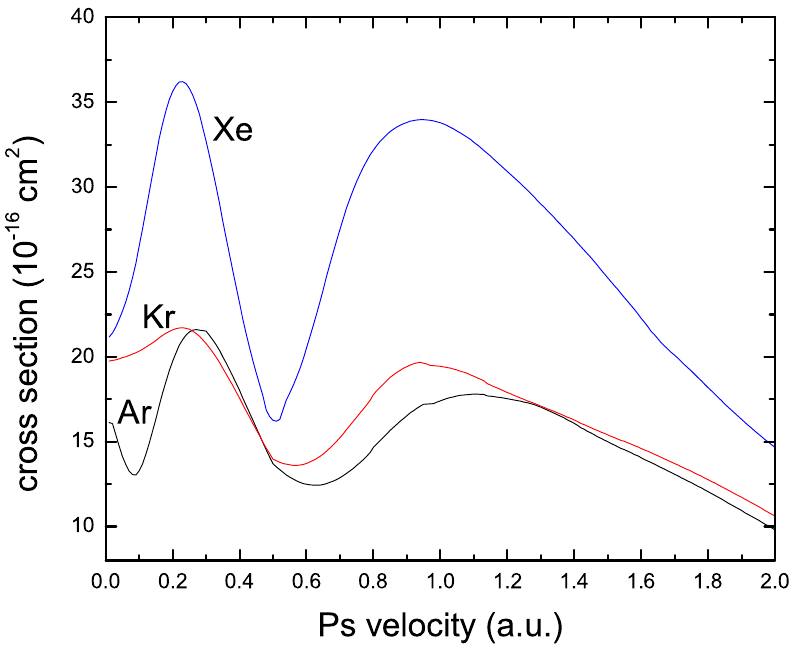}
\caption{Calculated Ps total scattering cross sections for the heavier rare-gas atoms. Data for Ar and Kr are taken from Ref.~\cite{Fab14b}.
}
\label{fig:Ps-rare}
\end{figure}
 
Figure \ref{fig:Ps-rare} presents a comparison of the calculated Ps total cross sections for all heavier rare-gas atoms. They exhibit the same features: the ``anti-Ramsauer'' maximum 
at low velocities and a broader maximum in the region $v\approx 1$~a.u. It is interesting that in a weaker form this feature is also present in the close-coupling Ps-hydrogen cross section when excitations of both Ps and the target (which account for the van der Waals interaction) are included (Fig.~5 in Ref.~\cite{Bla02}). Compared with Ar and Kr,
the absolute magnitude of the Xe cross section is substantially higher.
Note also that the Ar cross section is initially slightly decreasing,
indicating a relatively weaker role of the Pauli repulsion in this case.
However, due to the positive sign of the scattering length, the shallow minimum at $v=0.085$~a.u. is not of the same origin as the true RT minimum.

\begin{figure}
\includegraphics[width=\columnwidth]{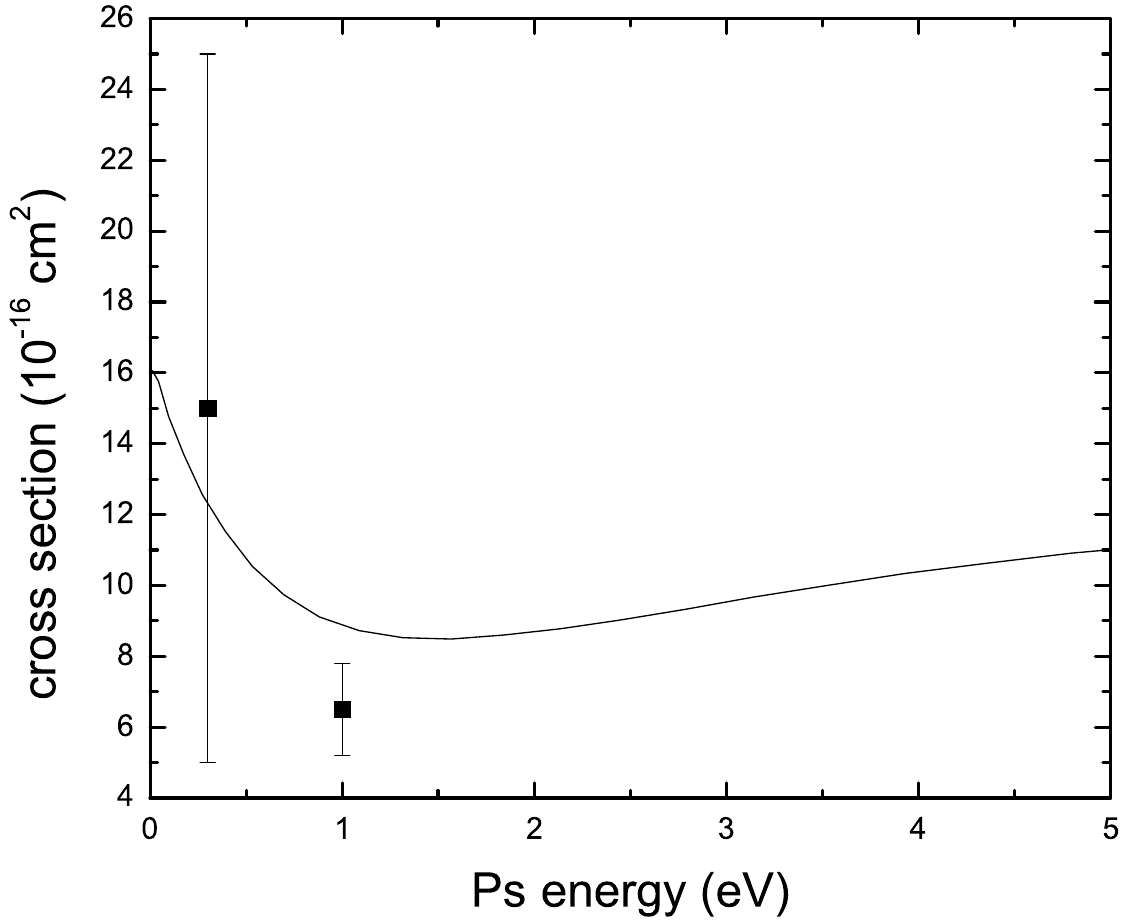}
\caption{Momentum-transfer cross section for Ps collisions with Ar at low energies: solid line is the calculation from Ref. \cite{Fab14b}; solid squares with error bars are the experimental data from Ref. \cite{Nag95} ($E=0.3$ eV) and Ref. \cite{Ska98} ($E=1$~eV).
}
\label{fig:mt}
\end{figure}

In Fig. \ref{fig:mt} we present the momentum-transfer cross sections for Ar in 
the low-energy region relevant to experiments \cite{Nag95,Ska98}
on Ps thermalization in Ar. Although the experimental error bars are large, 
the results of measurements are not inconsistent with the the tendency of the cross section 
to decrease in the energy range between 0 and 1 eV.
Our cross sections are also compatible with the measurements of Coleman 
{\it et al.} \cite{Col94}, who obtained the mean value $6.4\times 10^{-16}$ cm$^2$ 
in the energy
range between 0 and 6.8 eV, although it seems that our values are 
somewhat too high in this energy range. The fact that the cross sections
are likely
smaller in this region is also confirmed by the recent beam
measurements of Brawley {\it et al.} \cite{Bra15}.
   
\section{Binary-encounter approximation for P\lowercase{s} ionization}\label{sec:ioniz}

Ps ionization is one of the two main scattering channels in Ps collisions with neutral targets. At low collision energies comparable to the ionization threshold, the elastic scattering dominates, but for
energies of about 100~eV the ionization cross section becomes comparable 
to or even greater than the elastic cross section. Calculations show that Ps ionization is the major inelastic channel in Ps collisions \cite{Bla02a,Fab14a}. Hence, the sum of the elastic and Ps ionization cross sections provides a good approximation for the total cross section.

Owing to the diffuse nature of the Ps atom, the cross section of its ionization (or break-up) in collisions with atoms and molecules can be calculated using the impulse approximation \cite{Che52,Sta05}. There are two difficulties associated with such calculations. One is related to the ambiguity of the on-shell reduction of the electron and positron scattering amplitudes. 
The electron and positron scattering amplitudes which contribute to the
total amplitude in the impulse approximation are ``off the energy shell'', since they need to be evaluated for unequal initial ($v_i$) and final ($v_f$) electron or positron velocities, and energy $E\neq v_{i,f}^2/2$.
 The on-shell reduction of Starrett {\it et al.} \cite{Sta05}, which
was also used in Ref. \cite{Fab14a}, assumes that the amplitude is a function of the effective velocity $v=\max(v_i,v_f)$ and the momentum transfer $q$ is linked to the electron (positron) scattering angle $\theta_{\rm sc}$ by $q=2v\sin(\theta_{\rm sc}/2)$. In principle, one
can consider different methods of on-shell reduction, leading to different results for the Ps scattering amplitude and cross section.
The other difficulty is computational. In order to avoid lengthy calculations, Starrett {\it et al.} \cite{Sta05} used the so-called peaking approximation, which assumes that the Ps wave function in momentum space varies much faster than the scattering amplitude. While this assumption is justified, it introduces additional uncertainty in the results.

In this section we show that one can use a similar but simpler approach based on the binary-encounter approximation \cite{Smi76,Fla83} and avoid both difficulties in the calculations of Ps ionization. This approach employs the differential \textit{cross sections} rather than amplitudes, for electron and positron scattering from the target atom. 

Consider the process of Ps break-up in collision with a neutral target $B$: 
\[
B+{\rm Ps}\to B + e^+ + e^-.
\]
Assuming that at the instant of collision the electron and positron inside the Ps atom are quasi-free, the ionization rate due to $e^-$-$B$ collisions is \cite{Smi76}
\begin{equation} 
P_{\rm ion}=N_B\left\langle |{\bf v}-{\bf v}_B|\int_{\Delta E>I}d\sigma\right\rangle
\label{Smirnov1}
\end{equation}
where $N_B$ is the number density of particles $B$, ${\bf v}_B$ is the relative collision velocity, ${\bf v}$ is the
electron velocity relative to the Ps center of mass, $d\sigma$ is the
differential cross section for $e^-$-$B$ elastic scattering, and the integration is restricted to the angles which result in the energy transfer to the electron $\Delta E$ greater than the Ps ionization potential $I=6.8$~eV. The averaging denoted by $\langle \cdots \rangle$ is over the electron velocity distribution in Ps. A similar expression can be written for the $e^+$ contribution, and the total ionization rate is found by adding the two contributions and neglecting the interference between them.

Dividing Eq.~(\ref{Smirnov1}) by the flux density of incident particles $B$, we obtain the total ionization cross section due to electron interaction 
with $B$ as
\begin{equation}\label{ion}
 \sigma_{\rm ion}=\frac{1}{v_B}\left\langle |{\bf v}-{\bf v}_B|
\int_{\Delta E>I}d\sigma\right\rangle.
\end{equation}
In the laboratory reference frame the heavy particle $B$ is at rest, and
as a result of scattering, the electron velocity changes from ${\bf u}={\bf v}-{\bf v}_B$ to ${\bf u}'$, 
$|{\bf u}'|=|{\bf u}|$. The change of the electron kinetic energy in the
Ps frame then is
\[
\Delta E=\frac{1}{2}\left[\lvert{\bf u}'+{\bf v}_B\rvert^2-\lvert{\bf u}+{\bf v}_B\rvert^2\right]
={\bf v}_B\cdot({\bf u}'-{\bf u}).
\]
If we direct ${\bf v}_B$ along the $z$ axis and introduce spherical angles
$(\theta,\phi)$ and $(\theta',\phi')$ for the vectors {\bf u} and ${\bf u}'$, we obtain
\begin{equation}\label{eq:DelE}
\Delta E=v_Bu(\cos\theta'-\cos\theta).
\end{equation}
For the ionization process, integration over $\theta'$ is subject to the 
restriction
\begin{equation}\label{eq:constr}
I<\Delta E<v_B^2,
\end{equation}
where the upper limit follows from the Ps kinetic energy in the laboratory frame, consistent with the threshold for the ionization process, $Mv_B^2/2>I$, $M=2$ being the Ps mass. With the help of Eq.~(\ref{eq:DelE}), the constraints (\ref{eq:constr}) define the region in the $(\theta,\theta')$ plane:
\begin{equation}
 \cos\theta+\frac{I}{v_Bu}<\cos\theta'<\cos\theta+\frac{v_B}{u}. 
\label{constraints}
\end{equation}

The electron differential scattering cross section from a spherically symmetric target $B$ is
\[
\frac{d\sigma}{d\Omega}=\sum_{ll'}(2l+1)(2l'+1)f_{l'}^*f_lP_{l'}
(\cos\theta_s)P_l(\cos\theta_s),
\]
where $\theta_s$ is the scattering angle in the laboratory frame, i.e., the angle between 
{\bf u} and ${\bf u}'$, and $f_l$ is the scattering amplitude for partial wave $l$,
\[
f_l=\frac{1-e^{2i\delta_l(u)}}{2iu},
\]
defined by the phase shift $\delta_l(u)$. According to Eq.~(\ref{ion}), 
the differential cross section should be multiplied by $|{\bf v}-{\bf v}_B|=u$, integrated over the scattering angles, and averaged over the electron velocity distribution in the ground-state Ps,
\[
\frac{1}{4\pi}\left\lvert g_{1s}(v^2)\right\rvert^2=\frac{1}{4\pi}\frac{256}{\pi(4v^2+1)^4}.
\]
This is a five-dimensional integral with respect to the variables $\theta$, $\phi $, $\theta'$, $\phi'$, and $u$. Using the addition theorem for the spherical harmonics and writing
\[
Y_{lm}(\boldsymbol{\hat{\mathbf{u}}})=\Theta_{lm}(\cos\theta)\frac{e^{im\phi}}{\sqrt{2\pi}}, \]
where $\Theta_{lm}(\cos\theta)$ are the normalized associated Legendre functions, we can perform integration over the azimuthal angles $\phi$ and $\phi'$ with the result
\begin{align}
\sigma_{\rm ion}&=\frac{4\pi}{v_B}\int  _{I/2v_B}^{\infty}du\, u^3 \nonumber\\
&\times \int _{-1}^{1-I/v_Bu}d(\cos\theta)\left\lvert g_{1s}(u^2+v_B^2+2uv_B\cos\theta)\right\rvert^2 
{\nonumber}\\
&\times \sum_{ll'm}f_{l'}^*(u)f_l(u)\Theta_{l'm}(\cos\theta)\Theta_{lm}(\cos\theta) \nonumber \\
& \times \int_{\cos\theta+I/v_Bu}^{\cos\theta+v_B/u} d(\cos\theta')\Theta_{lm}(\cos\theta')
\Theta_{l'm}(\cos\theta').
\end{align}
The integration limits above follow from the restrictions (\ref{constraints}). The positron contribution has a similar form, with $f_l(u)$ being the positron scattering amplitudes from $B$.

\begin{figure}
\includegraphics[width=\columnwidth]{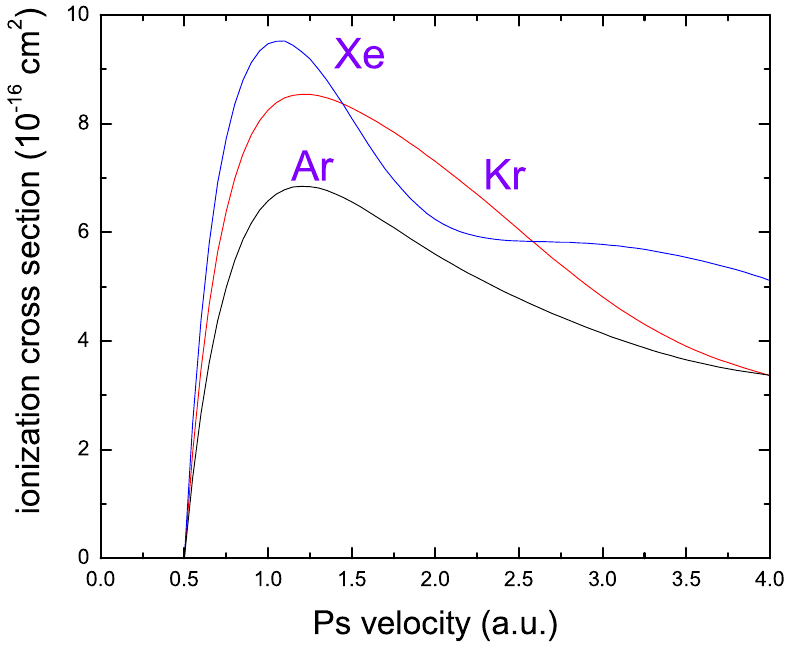}
\caption{Ionization cross section for Ps collisions with
heavier rare-gas atoms calculated in the binary-encounter approximation using the electron and positron phase shifts from the polarized-orbital method \cite{McE79,McE80,McE83,McE84}.
}
\label{fig:ionization1}
\end{figure}

To determine the Ps ionization cross sections for Ar, Kr, and Xe, we employed the electron and positron phase shifts calculated by McEachran and Stauffer using the polarized-orbital approximation \cite{McE79,McE80,McE83,McE84}. Figure \ref{fig:ionization1} shows the ionization cross sections for Ps collisions with Ar, Kr, and Xe over the range of velocities corresponding to energies from threshold to 435~eV. All the cross sections peak at a relatively low velocity, slightly above 1~a.u., and then decrease rather slowly for Ar and Kr, while for Xe the cross section drops and then has a broad second maximum at $v\approx 3$~a.u. 

\begin{figure}[ht!]
\includegraphics*[width=\columnwidth]{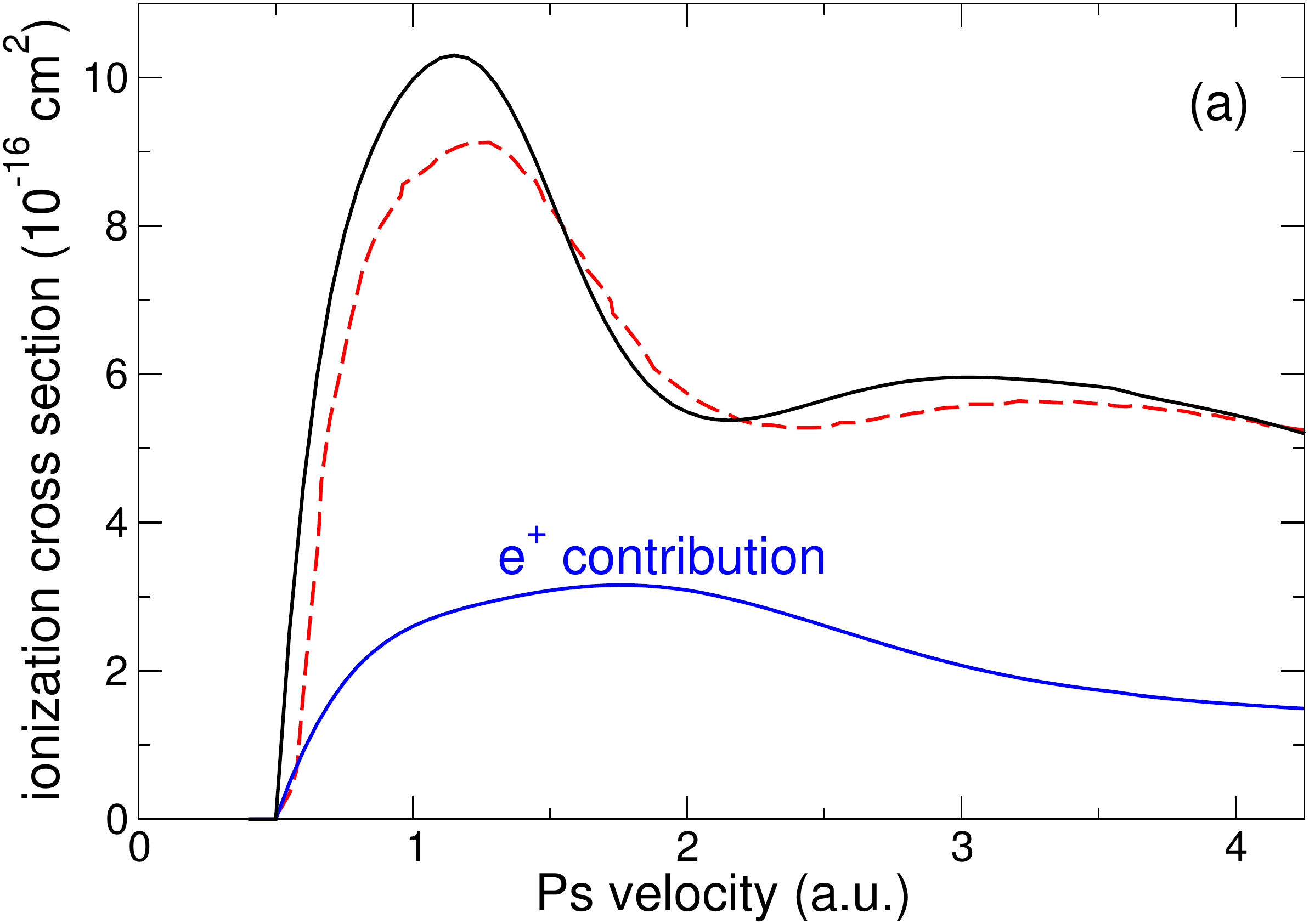}\\
\includegraphics*[width=\columnwidth]{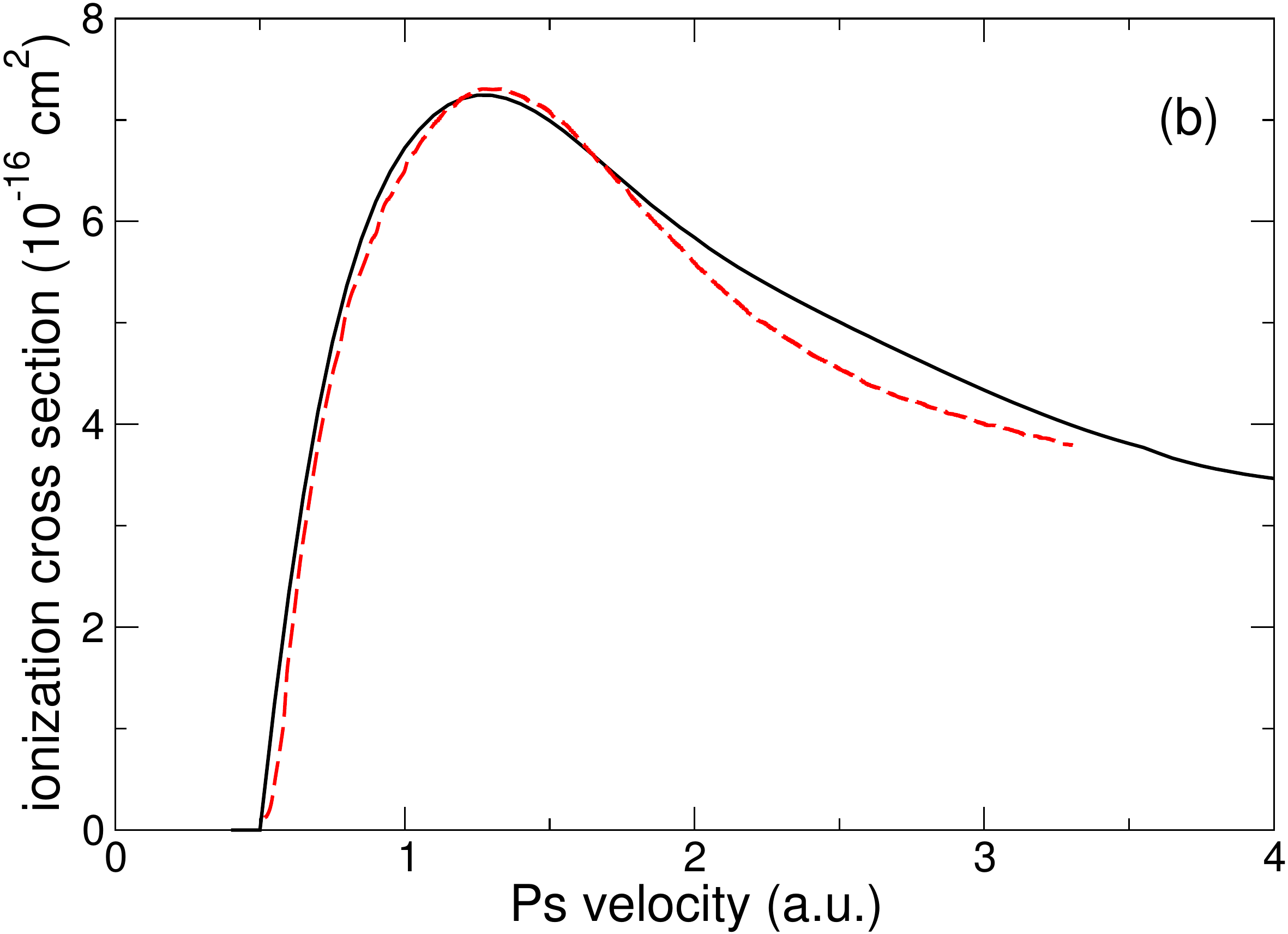}\\
\includegraphics*[width=\columnwidth]{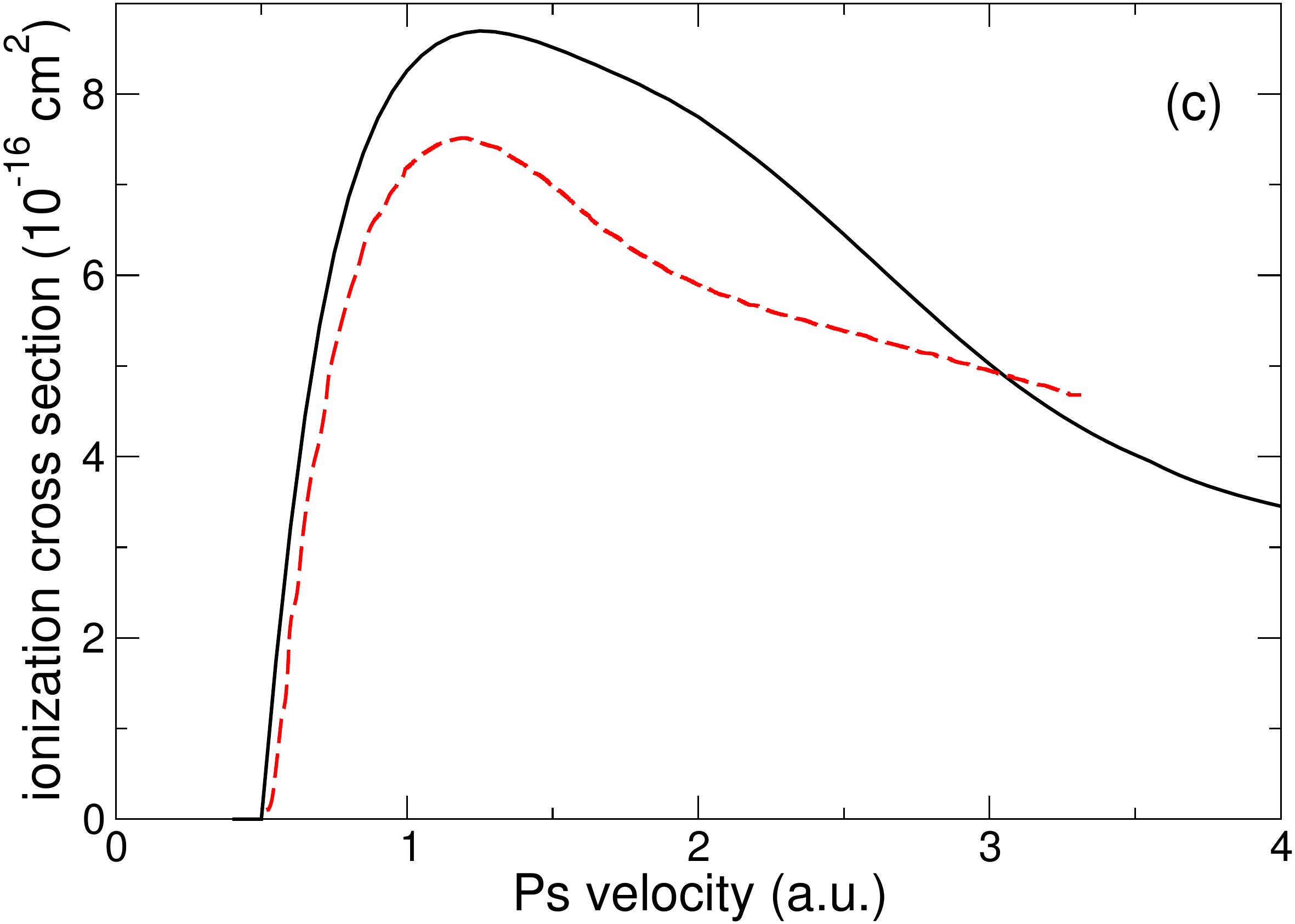}
\caption{Ionization cross section for Ps collisions with
(a) Xe, (b) Ar, and (c) Kr. Black solid line, present calculations; red dashed line, 
impulse approximation calculations of Starrett {\it et al.} \cite{Sta05}; blue solid line, the $e^+$ contribution to the ionization cross section (shown for Xe only).
}
\label{fig:ionization2}
\end{figure}

Figure~\ref{fig:ionization2} compares the present Ps ionization cross sections 
for Xe, Ar, and Kr with those obtained by Starrett {\it el al.} \cite{Sta05} 
using the impulse approximation. The agreement is generally good, especially 
for Ar. Note that the impulse approximation cross sections for Ar and Kr both 
peak at approximately $7.5\times10^{-16}~{\rm cm}^2$, while for Xe it peaks at 
approximately $10\times10^{-16}~{\rm cm}^2$. On the other hand, the 
binary-encounter cross sections show a progressive increase of the maximum for 
Ar, Kr, and Xe, which appears to be physically reasonable:
for heavier atoms the electron elastic scattering cross sections are higher, which
should lead to higher Ps ionization cross sections.
It is interesting that for Xe both the present binary-encounter approximation 
and the impulse approximation cross sections have the second maximum. Its 
position and magnitude differ slightly between the two calculations.

There might be several reasons for the discrepancies observed in Fig.~\ref{fig:ionization2}.
\begin{enumerate}
\item The electron and positron scattering amplitudes used by Starrett {\it et al.}~\cite{Sta05} were calculated used the static-exchange approximation, while the polarized-orbital phase shifts of McEachran and Stauffer used in the present work take target polarization into account.
\item The impulse approximation calculations of Starrett \textit{et al.} \cite{Sta05} uses on-shell reduction of the scattering amplitude, which is not a unique procedure. At the same time, the impulse approximation takes into account the interference between the electron and positron contributions, which is
neglected by the binary-encounter method used in the present work.
\item Starrett {\it et al.} \cite{Sta05} used the peaking approximation, which neglects the
velocity dependence of the scattering amplitude on the scale of the velocity 
spread of the Ps wave function in momentum space. This approximation might 
become less reliable at higher energies. On the other hand, the impulse 
approximation takes into account the Coulomb interaction within the 
electron-positron pair in the final state, while the binary-encounter 
neglects it.
\end{enumerate}
In view of all these different approximations made in the two methods, the 
agreement observed in Fig.~\ref{fig:ionization2} looks very satisfactory.

Experimental data for Ps collisions with Ar, Kr, and Xe atoms \cite{Bra12} do not indicate a second maximum or plateau in the total cross section as a function of Ps velocity. However, the measurements do not go above $v=2$~a.u., i.e., they perhaps have not reached the regime where the ionization cross section dominates the total. Note also that the velocity dependence of the measured total Ps-He and Ps-H$_2$ cross sections \cite{Gar96} becomes quite flat for velocities between 1.5 and 2 a.u.

\section{Conclusions}
Interaction of Ps with atoms is mostly controlled by the exchange interaction between the electron in Ps and the target electrons, and by the van der Waals interaction. For collision energies above the Ps ionization threshold the exchange interaction dominates, making the Ps-atom cross section look like $e^-$-atom cross section when plotted as a function of the
projectile velocity. These features are described very well by the present approach which combines the pseudopotential method for elastic Ps scattering and the binary-encounter approximation for Ps ionization (break-up). New calculations for Xe confirm the experimental observations \cite{Bra10,Bra10a,Bra12} of similarity between Ps-atom and electron-atom scattering. On the other hand, recent measurements \cite{Bra15} for Ar and Xe at low velocities do not confirm our predictions of the low-energy peak in the Ps-rare-gas-atom cross sections. It is possible that the pseudopotential model overestimates the $P$-wave contribution at low energies, and more theoretical work is necessary to describe Ps scattering in this region accurately.

\section*{Acknowledgments}
The authors are grateful to G. Laricchia for stimulating discussions
and for communicating Ps-Xe scattering data in advance of publication.
This work was partly supported by the US National Science Foundation 
under Grant No. PHY-1401788.

\bibliography{Ps-rare-gases}

\end{document}